\documentclass[namedreferences]{SolarPhysics}
\usepackage[optionalrh]{spr-sola-addons} % For Solar Physics
\usepackage{graphicx}        % For eps figures, newer & more powerfull
\usepackage{color}           % For color text: \color command
\usepackage{url}             % For breaking URLs easily trough lines
\newcommand{\etal}{{\it et al.}}

\newcommand{\adv}{    {\it Adv. Space Res.}}

\newcommand{\aap}{    {\it Astron. Astrophys.}}

\newcommand{\apj}{    {\it Astrophys. J.}}

\newcommand{\jgr}{    {\it J. Geophys. Res.}}

\newcommand{\pasj}{   {\it Pub. Astron. Soc. Japan}}

\newcommand{\solphys}{{\it Solar Phys.}}

\newcommand{\ssr}{    {\it Space Sci. Rev.}}

\begin{document}

\begin{article}

\begin{opening}

\title{The connection of solar wind parameters with microwave and UV emission of coronal hole atmosphere\\ {\it Solar Physics}}

\author{D.V.~\surname{Prosovetsky}$^{1}$\sep
        I.N.~\surname{Myagkova}$^{2}$
       }
\runningauthor{Prosovetsky and Myagkova} \runningtitle{The
connection of solar wind parameters...}

   \institute{$^{1}$ The Institute of Solar-Terrestrial Physics SB RAS, Irkutsk
                     email: \url{proso@iszf.irk.ru}\\
              $^{2}$ Lomonosov Moscow State University Skobeltsyn Institute of Nuclear Physics, MSU SINP, Moscow, Russia
              \\
                     email: \url{irina@srd.sinp.msu.ru} \\
             }

\begin{abstract}
This paper presents results of comparison between observations of
coronal holes in the UV (SOHO EIT) and microwave emission (17, 5.7
GHz, 327 and 150.9 MHz, NoRH, SSRT and Nancy radioheliographs), and
solar wind parameters, according to the ACE spacecraft data over the
period 12~March--31~May~2007. Increase in the solar wind velocity up
to $\sim$600 $km \cdot s^{-1}$ was found to correlate with decrease
in the UV flux in the central parts of the solar disk. The
connection between parameters of the microwave emission at three
different solar atmosphere levels and the solar wind velocity  near
the Earth's orbit was determined. This connection suggests existence
of common mechanisms of  solar wind acceleration from chromospheric
altitudes to the upper corona. We also suppose existence of two
different mechanisms of the solar wind acceleration at altitudes of
less and more than one solar radius.
\end{abstract}
\end{opening}
%-------------------------------------------------

\section{Introduction}
     \label{S-Introduction}

The solar wind (SW), predicted by Vsechsvyatsky
\cite{Vsehvyatskiy55}, Ponomarev \cite{Ponomarev57}, Parker
\cite{Parker58}, was found in the experiments on board the
spacecrafts Luna-2 and Luna-3 \cite{Gringauz62}. Solar wind can be
sorted out into three components -- the slow, the fast, and the
sporadic ones.  Existence of different components of the solar wind
has been revealed due to the observations of the spacecraft
Mariner-2 \cite{Neugebauer66}. According to \opencite{Sheeley85},
the sporadic component is connected with coronal mass ejections.
Unlike the sporadic component, the slow and fast components are
regular: the slow one exists always, and the fast one is observed
periodically. Sources of the latter are coronal holes (CH) --
regions of the unipolar magnetic field with open configuration
\cite{Hundhausen72}.

The slow and fast SW components differ not only in plasma velocity
(up to ~450 and ~800 $km \cdot s^{-1}$, respectively). According to
the observations of radio scintillations on inhomogeneities of the
interplanetary medium, one of the main SW parameters - velocity
distribution depending on the distance from the Sun - is different
for the slow and fast components. The maximum velocity of the slow
component is reached at a distance of more than 10 solar radii
\cite{Wang98}, whereas the fast component reaches its maximum
velocity near the Sun, at a distance of one or two solar radii
\cite{Grall96}. These results of observations suggest that the
mechanism accelerating SW particles is either completely different
or has essentially different parameters for different components.
Differences in conditions of the fast SW component acceleration are
likely to be found at distances of less than one solar radius over
the photosphere.

This assumption is also supported by observations of the velocity
nonthermal component made with the use of UV spectrograph SUMER on
board SOHO \cite{Wilhelm95}. Investigation into ultraviolet spectrum
of the CH atmosphere \cite{Chae98} shows that the maximum of
fluctuation velocity (i.e., the nonthermal component $\delta V$ in
the expression $V_{solar}=\sqrt{\frac{2kT}{M}+\delta V^2}$, where
$k$ is the Boltzmann constant, $T$ is the ion temperature, $M$ is
the mass of the ion emitting the line) is reached  at CH levels in
the solar atmosphere with temperatures of about $10^5$ K (in the
transition region). The nonthermal velocity component specifies wave
flux $F_{wave} = \rho \langle \delta V^2 \rangle V_A$, where $\rho$
is the plasma density, $V_A$ is the Alfven velocity that is probably
responsible for acceleration of SW particles and heating of the
corona into CH.

Consequently, parameters of the fast SW component may be formed in
the lower solar atmosphere (chromosphere, transition region, and
lower corona). Some authors also confirm that SW parameters and the
solar emission in the optical and UV waveband are related (e.g.,
\opencite{Vrsnak07}; \opencite{Obridko09}). Intensity, areas and
ratio of areas of CH emission/absorption lines were proposed  in
paper \cite{Stepanian08, Shugai09, Obridko09} to predict SW
characteristics. However, these methods are generally based on the
emission analysis at single wavelength (sometimes at two
wavelengths); i.e., they do not reflect completely distribution of
plasma parameters and energy release processes in the solar
atmosphere.

Supplementary information about peculiarities of acceleration of the
fast SW component could be gained from the thorough study of the UV
and microwave emission of the CH, obtained with high spatial
resolution. However, there are only a few works (e.g.,
\opencite{Chae98}) devoted to spectral observations of the UV
emission obtained from SUMER data. This is caused by the
peculiarities of observation programmes of this instrument. Besides,
no studies of dependence between high-speed SW characteristic and
the microwave CH emission have been made so far.

Different scientific groups have earlier made observations of the
microwave CH emission over a wide frequency range. Comparative
analysis of such observations shows that the frequency range
6-17~GHz is characterized by an increased emission (compared to the
quiet Sun). This fact can not be explained in the framework of
typical models of the solar atmosphere \cite{Maksimov06}. The
Nobeyama Radioheliograph (NoRH, 17~GHz, \opencite{Nakajima94})
regularly observes the increased level of brightness temperatures in
CH \cite{Gopalswamy99, Nindos99, Moran01}. Analysis of simultaneous
observations with NoRH and the Siberian Solar Radio Telescope (SSRT,
5.7~GHz, \opencite{Grechnev03}) revealed linear anticorrelation
between brightness temperatures at 5.7 and 17~GHz in coincident CH
regions \cite{Krissinel00}. According to the results of observations
near the solar limb, carried out at SSRT and NoRH, such regions are
situated radially \cite{Maksimov04}. Analysing observations of the
microwave emission, \cite{Maksimov06} showed that the corona (at
least, in some parts of CH) is heated by the wave flux propagating
from the photosphere, and the altitudes where the increased
microwave emission is formed correspond to the altitudes of the
chromosphere and transition region. Some researchers (see
\opencite{Cranmer04}) suppose the same mechanisms of the SW
acceleration and coronal heating. Observations of the UV spectrum
imply possible connection between energy release and microwave
emission in the lower solar atmosphere.

According to the study of the magnetic field configuration of CH,
there is a connection between SW characteristics and parameter of
the "superradial" divergence of the magnetic tube. "Superradial"
divergence is determined from measurements of the photospheric
magnetic field and its extrapolation (in the potential
approximation) to the source surface \cite{Wang90}. There is also a
connection between SW characteristics and the unipolar field area at
the CH base \cite{Eselevich09}. Forecast of SW parameters based on
such works does not always yield correct results . Probably the
magnetic field configuration can not completely determine
concentration and velocity of SW, since the SW plasma carries
frozen-in magnetic flux out to the outer solar atmosphere and thus
changes its configuration. This restricts application of methods of
the magnetic field extrapolation (potential and non-potential) which
do not take this effect into account when forecasting SW parameters.

The goal of this investigation is to study possible connection
between the SW velocity and emission of the lower solar atmosphere
of CH.

\section{Observations} %%%%%%%%%%%%%%%%%%%%%%%%%%%%%%%%%%%%%%%%
      \label{S-Observations}

\subsection{Data} %%%%%%%%%%%%%%
  \label{S-text}
The period from 12 March to 31 May 2007 (the last solar minimum) was
studied. We chose this interval because of the fact that the main
parameters of the fast SW component during a solar minimum are
defined by CH characteristics. Besides, there was a complete set of
experimental data for this period.

To perform analysis, we used images of the Sun in the microwave
emission recorded by NoRH, SSRT and Nan\c{c}y radioheliographs at
17~GHz, 5.7~GHz, 327~MHz and 150~MHz, and the SOHO/EIT data at the
wavelength Fe XII $\lambda$=195\AA. Variations in SW parameters were
measured by SWEPAM (Solar Wind Electron, Proton, and Alpha Monitor)
during the experiment on board the ACE spacecraft (Advanced
Composition Explorer) orbiting near the libration point L1 in the
Sun-Earth system (1.5 million km from the Earth towards the Sun),
see \url{http://cdaweb.gsfc.nasa.gov/istp_public}.

\subsection{Data processing} %%%%%%%%%%%%%%
Applying the method of separation of coronal hole boundaries similar
to the technique that was used in \cite{Vrsnak07}, we found integral
characteristics with the use of two-dimensional images of the Sun.
This method implies separation of a region bounded by meridians near
the solar centre. Then, the UV emission flux was calculated in this
region. We noticed, however, that the polar regions did not
contribute significantly to the SW stream registered near the Earth.
Thus the region (unlike that in \opencite{Vrsnak07}) was bounded by
an ellipse $R^2_{\bigodot}=X^2+\frac{Y^2}{1.9}$, where
$R_{\bigodot}$ is the solar radius. Relation between semiaxes of the
ellipse was determined when comparing positions of CH boundaries and
the SW velocity at the point L1 (i.e., we were trying to find out if
the high-speed SW stream was registered from CH that had the
latitude and longitude corresponding to the occurrence of a CH
boundary on ellipse boundaries).

The SOHO/EIT data were processed using \textit{SolarSoft} procedures
and images from the calibration database of the instrument. The
ratio of the emission flux to the CH area (i.e., the average
emission flux in the Fe~XII~$\lambda$=195\AA~line) was determined
for CH within the limits of an elliptic central region. The ionic
concentration $N_i$ in CH was deduced from the relationship:
   \begin{equation}  \label{Eq-EIT2den}
   N_i=8.34+0.509\cdot \log(I_{EIT})
   \end{equation}
obtained in \opencite{Brosius02}, where $I_{EIT}$ is the
Fe~XII~$\lambda$=195\AA~line intensity on the calibrated image.

Using data on the microwave emission at 4 frequencies, we determined
equivalent to the microwave flux of CH for the regions inside the
ellipse, in the quiet Sun units:
 \begin{equation}  \label{Eq-MW-flux}
   F_{\rm CH,\nu} = \frac{\sum \limits _{\rm i=1}^{\rm n_{CH}}(T_{\rm B\nu,i}-T_{\rm QS\nu,i})}{\sum \limits _{\rm j=1}^{\rm n}T_{\rm QS\nu,j}}
 \end{equation}
where $\nu$ is the frequency, $T_{\rm B\nu,i}$ is the brightness
temperature inside a CH (its summation is made throughout the CH
surface elements $S_i$), $T_{\rm QS\nu,i}$ is the brightness
temperature of the quiet Sun (its summation is made throughout the
entire disk). So we obtained data on the microwave emission of CH
from 4 atmospheric levels, and data on the ionic concentration
inside CH.

\section{Experimental data analysis} %%%%%%%%%%%%%%%%%%%%%%%%%%%%%%%%%%%%%%%%
      \label{S-Analysis}

\subsection{Solar wind parameters} %%%%%%%%%%%%%%

Three upper panels (a, b, c) of Figure 1 show SW parameter
variations during the period under study: strength of the $B_z$
component and $B$ of the interplanetary magnetic field (IMF), the SW
velocity, $V_{SW}$ and the proton concentration $N_p$ in SW,
respectively. Referring to Figure 1, ten high-speed SW streams
(whose beginning is marked by vertical lines) are evident throughout
this period. The first SW stream (with its maximum on 13 March) was
excluded from consideration, since this SW stream was generated by
CH that had passed through the central meridian some days before 13
March (when the UV imaging telescope EIT started its observations
after regular maintenance service). Lack of EIT data (139-145 days
of year) is caused by the break in observations of the instrument
during these days. In Figure 1, the streams considered in the work
are numbered from 1 to 9.

 \begin{figure}    %%%%%%%%%%%%%%%%%% FIGURE 1
   \centerline{\includegraphics[width=0.7\textwidth,clip=]{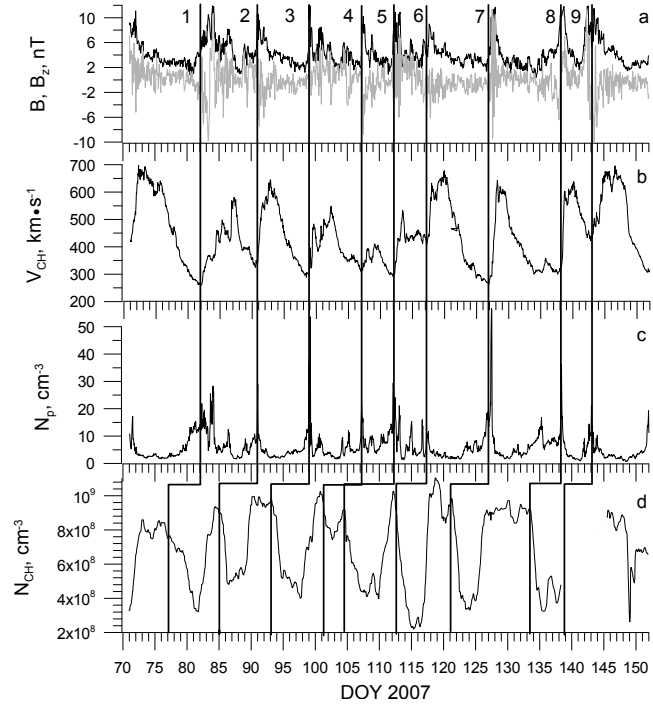}
              }
              \caption{
SW characteristics measured at the point L1 on 12 March - 31 May
2007: modulus B (black line) and the Bz component (grey line) of IMF
(a); the SW velocity VSW (b); the proton concentration Np (c). The
panel "d" shows variation of the plasma concentration in CH NCH
(curve discontinuity corresponds to the break in SOHO EIT
observations). The x-coordinate is the number of days, starting from
1 January (DOY). Vertical lines correspond to the leading edge of
the SW velocity variation and to the beginning of plasma
concentration decrease in CH below the background level.
                      }
   \label{F-1}
   \end{figure}

The high-speed streams observed during the period under study have
essentially different characteristics: the maximum speed varies from
300 to 680~$km \cdot s^{-1}$, the proton concentration at the
leading edge of the high speed stream can reach up to
40-45~$cm^{-3}$ or can be less than 5~$cm^{-3}$ (Figure 1b,c).
Variations in the average IMF value B for 9 SW streams look similar:
an abrupt increase at the leading edge of the SW stream and
subsequent gradual decrease. During the period under study, $B$
varied in the range from 2 to 10~nT.

\subsection{UV emission of CH} %%%%%%%%%%%%%%

The lower panel "d" of Figure 1 shows variation of plasma
concentration in CH.  It is observed from the figure that the main
increases in the SW velocity (1-9), taking account of the arrival
time at the point L1, coincide with plasma concentration decrease in
CH. Notice that the delays, determined from concentration fronts in
CH and from SW stream speeds, are different and vary from 4 to 7
days.

\begin{figure}    %%%%%%%%%%%%%%%%%% FIGURE 2
   \centerline{\includegraphics[width=1.0\textwidth,clip=]{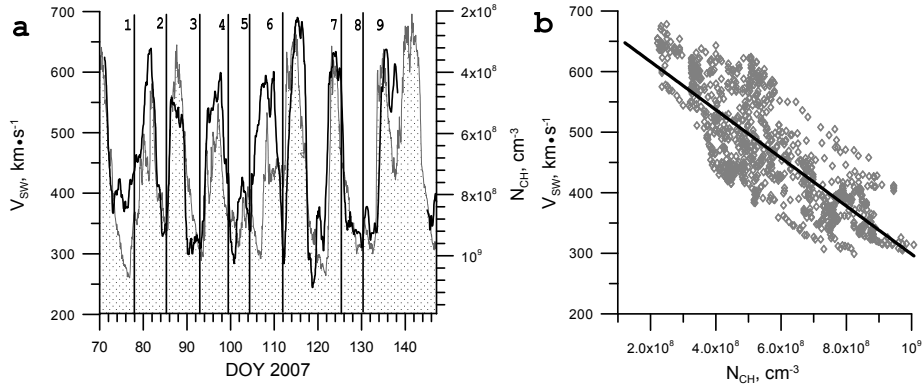}
              }
              \caption{
SW velocities at the point L1 and plasma concentration in CH.
?.~Superposed variations (shift is 4 days) of concentration (thick
line) and in the SW velocity (filled curve bounded by thin line).
The y-axis of the plasma concentration values is reversed. b.
Correlated dependence between concentration $N_{CH}$ and velocity
$V_{SW}$, with account taken of different delays in time of the SW
stream arrival at the point L1. The line shows best fit for the
points.                      }
   \label{F-2}
   \end{figure}

In Figure 2a, diagram of SW velocity variations in time is
superposed onto diagram of variations in the mean concentration of
CH, with consideration for the time delay equal to the time when SW
particles with propagation velocity 600 $km \cdot s^{-1}$ come to
the point L1. The y-axis for concentration (to the right) is
reversed (i.e., higher values are at the bottom). Referring to
Figure 2a, SW velocity variations and plasma concentration in CH are
similar. Discrepancy between positions of some SW streams and
concentration depressions (e.g., streams 4, 5, and 6) can be
explained by the difference in time of the SW particle arrival at
the point L1.

Taking account of different delays in the SW particle arrival (from
4 to 7 days, see Figure 1) at the point L1, we show correspondence
between concentrations and velocities $V_{SW}\geq$350 $km \cdot
s^{-1}$ (Figure~2b). The slow SW component is thus cast out in the
time series where possible. Values $N_{CH}$ and $V_{SW}$ in 8 SW
streams are approximated by the linear dependence
\begin{equation}  \label{Eq-EIT-N}
   V_{\rm SW} = -3.97 \cdot 10^{-7}N_{\rm CH}+695
\end{equation}
with correlation 0.63. Notice that parameters of different SW
streams significantly differ (especially the delay time between the
plasma concentration decrease in CH and the arrival of the SW
high-speed stream at the Earth). Correlation of each separate SW
stream velocity with its corresponding linear dependence reaches
0.8-0.9. However, there was no evidence of correlation between
plasma concentration in CH and proton concentration in SW.
 \begin{figure}    %%%%%%%%%%%%%%%%%% FIGURE 3
   \centerline{\includegraphics[width=0.8\textwidth,clip=]{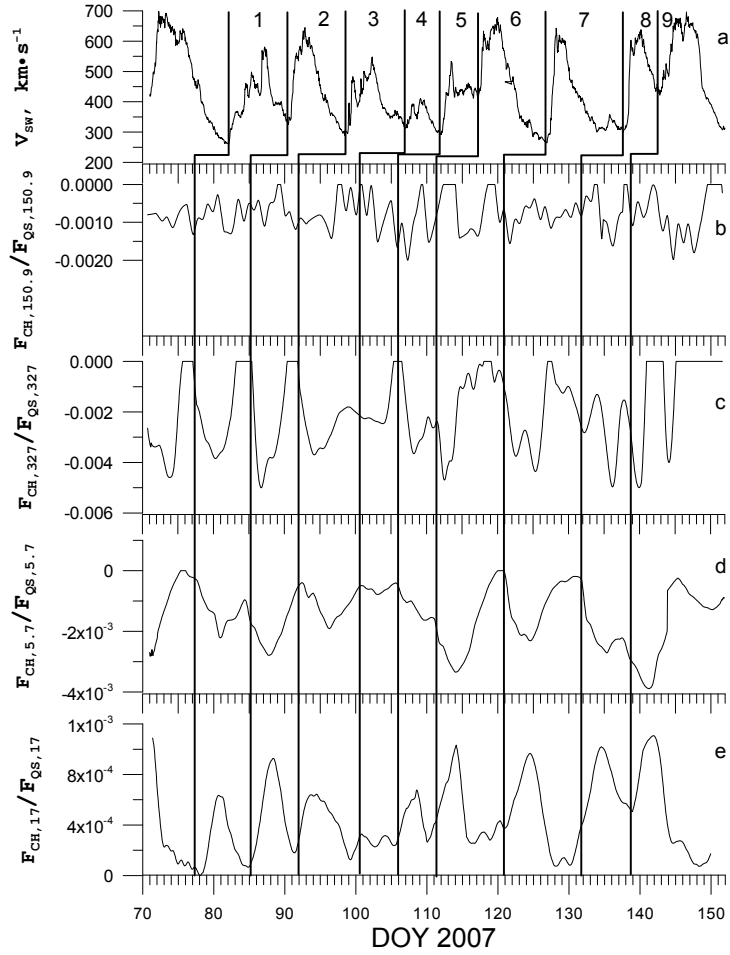}
              }
               \caption{
Microwave emission fluxes of CH at 150.9~MHz, 327~MHz, 5.7 and
17~GHz (b, c, d, e) in the quiet Sun units, compared to the SW
velocity (a). Vertical lines correspond to the leading edge of SW
streams, to the beginnings of CH flux depressions at 150.9~MHz,
327~MHz, 5.7 GHz, and of increases of the microwave flux at 17~GHz.
 }
   \label{F-3}
   \end{figure}

\subsection{Microwave emission of CH} %%%%%%%%%%%%%%

Figure 3 (b-e) presents variations in the microwave emission flux of
CH at 4 frequencies, according to data from NoRH, SSRT and Nan\c{c}y
radioheliographs. The flux was measured with the use of the method
described in the Section 2. Vertical lines in Figure 3a are placed
on leading edges of SW streams; those in Figure 3 (b-e), on leading
edges of increments/decrements of microwave flux at 4 frequencies.
The increased microwave emission flux is related to CH at 17~GHz,
whereas the decreased microwave emission flux is related to it at
5.7~GHz and 327~MHz. At 150.9~MHz, there is no evidence of an
increased or decreased microwave emission flux. Its magnitude is
comparable to the flux of neighbouring regions of the solar
atmosphere (Figure 3b). Many authors \cite{Dulk74, Trottet78,
Kosugi86, Borovik90, Chiuderi-Drago99, Gopalswamy99, Nindos99,
Krissinel00, Moran01, Maksimov06} indicated the increased microwave
emission at frequencies near 17~GHz, the decreased microwave
emission at frequencies below 5~GHz and above 150~MHz, and the
drastic decrease in the CH contrast relative to the quiet Sun at
frequencies below 150~MHz.

Nine CH were registered during the period under study; time of the
SW stream arrival at the point L1 was taken into consideration
(Figure~3), and microwave emission fluxes with SW velocities
$V_{SW}$ $\geq$350~$km \cdot s^{-1}$ were compared at 4 frequencies
(Figure 4 a-c). The following dependences were obtained from the
linear approximation of experimental data:

   \begin{equation}  \label{Eq-MW-17}
   V_{\rm SW} = 3.04 \cdot 10^{5}F_{\rm CH, 17}+313
   \end{equation}
   \begin{equation}  \label{Eq-MW-5.7}
   V_{\rm SW} = -1.00 \cdot 10^{5}F_{\rm CH, 5.7}+300
   \end{equation}
   \begin{equation}  \label{Eq-MW-327}
   V_{\rm SW} = -7.85 \cdot 10^{4}F_{\rm CH, 327}+247
   \end{equation}

Observational data correspond to these dependences with correlation
0.69 at 17~GHz, 0.84 at 5.7~GHz, and 0.60 at 327~GHz. No reliable
dependence was found for 150.9 MHz.

  \begin{figure}    %%%%%%%%%%%%%%%%%% FIGURE 4
   \centerline{\includegraphics[width=1.0\textwidth,clip=]{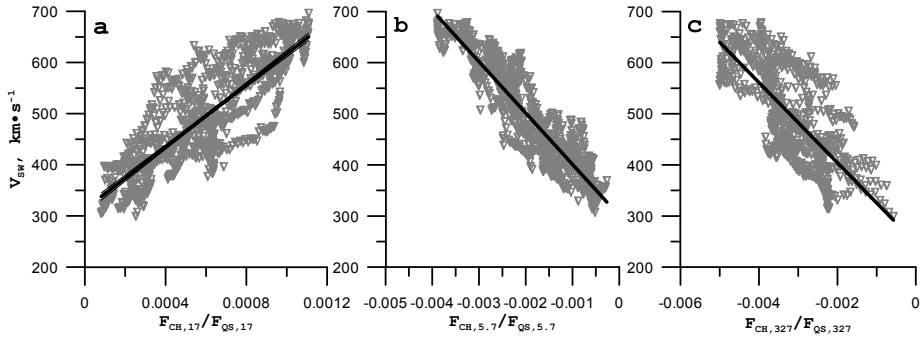}
              }
              \caption{
Correlated dependences between ratio of the CH microwave flux  and
quiet Sun flux at 17 (a), 5.7~GHz (b) and 327~MHz (c) and the SW
velocity at the point L1. Lines show the best fit for a
corresponding dataset.                      }
   \label{F-4}
   \end{figure}

\section{Discussion} %%%%%%%%%%%%%%%%%%%%%%%%%%%%%%%%%%%%%%%%

When analysing experimental data in the previous section, we also
examined probable connection between other SW parameters ($B$,
$B_Z$, $N_p$) and CH atmosphere characteristics (the microwave
emission flux $F$ at 4 frequencies, the UV intensity, and its
related ionic concentration $N_{CH}$). No reliable relations was
revealed. Probably these SW parameters are formed at other
atmospheric levels of CH or in interplanetary space.

Among SW parameters under study, we found dependences only between
its velocity at the point L1 and characteristics of the solar
atmosphere emission in UV and in the microwave at 3 frequencies.
Some authors (e.g., \opencite{Vrsnak07, Obridko09}) have already
called attention to the relation between SW stream speeds and some
characteristics of the UV emission (for instance, the area) in CH.
In this work, we established dependence of the SW velocity to the
mean plasma concentration in CH and the microwave flux at 3
frequencies.

\sloppy
According to the linear dependencies $V_{SW}(N_{CH})$
(\ref{Eq-EIT-N}), $V_{SW}(F_{CH,~17})$ (\ref{Eq-MW-17}),
$V_{SW}(F_{CH,~5.7})$ (\ref{Eq-MW-5.7}), and $V_{SW}(F_{CH,~327})$
(\ref{Eq-MW-327}), all CH characteristics ($N_{CH}$, $F_{CH,~\nu}$),
and the SW velocity $V_{SW}$ are interrelated. Actually, the
dependencies $N_{CH}(V_{SW})$, $F_{CH,~17}(V_{SW})$,
$F_{CH,~5.2}(V_{SW})$ and $F_{CH,~327}(V_{SW})$ obtained from
(\ref{Eq-EIT-N}), (\ref{Eq-MW-17}), (\ref{Eq-MW-5.7}) and
(\ref{Eq-MW-327}) form a parametric equation of line:

   \begin{equation}  \label{Eq-arr}
   \left\{
\begin{array}{rcl}
N_{\rm CH}=-2.52 \cdot 10^6V_{\rm SW}+1.75 \cdot 10^9\\
F_{\rm CH, 17}=3.29 \cdot 10^{-6}V_{\rm SW}-1.03 \cdot 10^{-3}\\
F_{\rm CH, 5.7}=-1.00 \cdot 10^{-5}V_{\rm SW}+3.00 \cdot 10^{-3}\\
F_{\rm CH, 327}=-1.27 \cdot 10^{-5}V_{\rm SW}+3.15 \cdot 10^{-3}
\end{array}
\right.
   \end{equation}
\\
and all variables of this system are also linearly dependent. This
result confirms linear dependence between brightness temperatures at
5.7 and 17 GHz in CH, revealed by \opencite{Maksimov06}. Notice that
no dependence between the SW velocity and the microwave emission
flux was found for 150.9 MHz at 4 frequencies. The microwave
emission flux at this frequency is therefore linearly independent of
the emission at other 3 frequencies and of the plasma concentration
determined from the Fe~XII~$\lambda$=195\AA~line emission.

The solar atmosphere in quiet regions from the chromosphere to the
corona is thought to be optically thin for the thermal free-free
microwave emission. In this case, the brightness temperature is
$T_B\approx\tau T_e$, where $\tau \ll 1$ is the optical thickness,
$T_e$ is the electron temperature. For the thermal free-free
emission:

   \begin{equation}  \label{Eq-MW-Thin-Flux}
   F_{\rm \nu} = \tau T_e \cdot S \cdot L \approx \frac{9.786 \cdot
   10^{-3}N^2}{\nu^2T_e^{\frac{1}{2}}}\ln\Lambda \cdot S \cdot L
   \end{equation}
where $F_{\rm \nu}$ is the microwave flux of CH at the frequency
$\nu$, $N$ is the electron concentration in CH, $S$ is the area of
the emitting region, L is the thickness of the emitting layer,
$\ln\Lambda$ the Coulomb logarithm.

The exact height of the Fe~XII~$\lambda$=195\AA~line formation in CH
is unknown. Though this line is considered to be coronal, CH limb
observations show (according to SOHO EIT data) that the line
intensity (and comparable values of plasma concentration in CH
within the range from $10^7 cm^{-3}$ to $10^9 cm^{-3}$) is observed
at heights less than 1 solar radius. This suggests that plasma
concentration values in CH obtained in this work correspond to range
of altitudes where the microwave emission is formed at 17~GHz
($\sim$3000 km), 5.7~GHz ($\sim$20~000 km), and 327~MHz
($\sim$100~000--200~000~km). So we can expect that there will be a
frequency or a range of frequencies whose formation altitudes would
correspond to the Fe~XII~$\lambda$=195\AA~line emission altitude,
and there may be a linear dependence for these frequencies between
the emission flux from CH and the SW velocity. We can also expect
that the height of 1 out of 3 frequencies considered in this work is
close to the Fe~XII~$\lambda$=195\AA~line formation altitude.

However, there is a contradiction between the revealed linear
dependence of the microwave emission flux (\ref{Eq-MW-Thin-Flux}) on
$N_{CH}$ and formula (\ref{Eq-arr}) where dependence is square-law
for the optically thin CH atmosphere. This may be explained in two
ways. The first explanation is that dependence  $F_\nu(N)$ is linear
for the obtained range of $N$.  However, When substituting the
concentration values measured in CH (see Figure 2b) into (8), we can
easily find that dependence $F_\nu(N)$ is not linear at 17, 5.7~GHz
and 327~MHz, within the temperature range from the chromospheric to
the coronal one. The second explanation is that there is a layer
with an increased energy release at heights from the chromosphere to
the corona. In this case, the atmosphere is optically thick for the
microwave emission, i.e., $T_B=T_e$, and the linear dependence
between the microwave flux and plasma concentration is generated by
energy release.

According to \cite{Aihua89}, such conditions correspond to the
regions with decreased plasma concentration and open magnetic field
- coronal holes, and energy is transferred by a wave flux, possibly
by Alfven waves. Observations of CH at 17~GHz \cite{Gopalswamy99,
Nindos99}, where the brightness temperature of CH is higher than
that of surrounding quiet regions, confirm energy release in the
upper chromosphere and transition region. Besides, observations of
CH at frequencies near 17~GHz also demonstrate an increased
brightness temperature in CH \cite{Gopalswamy99, Nindos99,
Krissinel00, Moran01, Maksimov06}. The linear dependence between
brightness temperatures at 5.7 and 17~GHz in CH \cite{Maksimov06},
on the assumption that there is a wave flux propagating from the
lower solar atmosphere and dissipating in the chromosphere and
transition region, explains relation between the increased emission
in the chromosphere and transition region and the simultaneous
cooling of the lower corona. If heating of the CH atmosphere is
caused by dissipation of the wave flux, the more wave flux is
dissipated in the lower solar atmosphere, the less energy is
released in the upper solar atmosphere.

The connection between SW parameters and microwave emission
established in this work provides a good reason to believe that
heating of the solar atmosphere and acceleration of the high-speed
SW are of similar nature. Some authors \cite{Hollweg78, Tu95, Tu05,
Suzuki06} think that acceleration of SW particles is caused by the
energy transfer from the wave flux to them (for instance, due to the
wave-particle resonance), and wave sources are in the lower solar
atmosphere. According to data from the SUMER instrument on board
SOHO, \cite{Chae98} found that the maximum value related to the wave
flux is in the transition region. Measured velocities of matter in
the polar CH are about 130-160~$km \cdot s^{-1}$ (already in the
chromosphere and transition region) and up to 330~$km \cdot s^{-1}$
in the lower corona, according to data from the instruments
EIS/Hinode and SUMER/SOHO \cite{Gupta10}. This implies that the
particle acceleration in the high-speed SW related to CH is observed
already in the lower solar atmosphere, and connection of the SW
velocity near the Earth's orbit with the emission at these heights
becomes clear.

Whatever SW acceleration mechanisms in CH are at altitudes of less
than 1 solar radius, they have little or no effect at heights of
more than 1 solar radius. This is confirmed by the absence of
connection between the microwave emission flux in CH at 150.9~MHz
and the velocity of the high-speed SW, whereas such a connection is
in evidence at higher frequencies. The works \opencite{Dulk74,
Chiuderi-Drago74, Trottet78, Chiuderi-Drago99} also show that CH at
$\sim$150 MHz and below may not be visible on the quiet Sun
background or slightly differ in brightness temperature from the
temperature of the quiet Sun. Measurements of plasma velocity with
UV coronal spectrographs may reveal whether SW particles at coronal
heights are accelerated or not. However, no works on this issue have
been written so far. We hope that they will appear due to data from
new space observatories (e.g., SDO).

\section{Summary} %%%%%%%%%%%%%%%%%%%%%%%%%%%%%%%%%%%%%%%%

From the analysis of available experimental data, the following
conclusions have been made:

1. Velocity of the high-speed SW during the period under study is
related to the mean emission intensity of the
Fe~XII~$\lambda$=195\AA~line (and the mean plasma concentration NCH
determined from it) in CH.

2. The connection has been established between velocity of the
high-speed SW and the CH microwave emission flux in the chromosphere
and lower corona during the period under investigation.

3. The evidence we have found indicates existence of the common
mechanism of the high-speed SW acceleration from the chromosphere to
corona; it also proves existence of two different mechanisms of the
high-speed SW acceleration at distances of less and more than one
solar radius.

4. We have obtained evidence that there are common mechanisms of the
coronal heating and the high-speed SW acceleration.

%%%%%%%%%%%%%%%%%%%%%%%%%%%%%%%%%%%%%%%%%%%%%%%%%%%%%%%%%%%%%%%%%%%%%%%%%%%
\begin{acks}
We are grateful to the team of observatories SOHO (instrument EIT),
SSRT, Nobeyama and Nancay for providing free access to data that
enabled this work.
\end{acks}

%%% BIBLIOGRAPHY %%%%%%%%%%%%%%%%%%%%%%%%%%%%%%%%%%%%%%%%%%%%%%%%%%%%%%%%%%%

\end{article}


\begin{thebibliography}{}

\bibitem[\protect\citeauthoryear{{Aihua \etal}}{1989}]{Aihua89}
Aihua,~Z., Daxiong,~F., Jianmin,~W., Chunmei,~L.: 1989, \adv,
\textbf{9}(4), 33.

\bibitem[\protect\citeauthoryear{{Borovik
\etal}}{1990}]{Borovik90} Borovik,~V.N., Kurbanov,~M.S.,
Livshits,~M.A., Ryabov,~B.I.: 1990, \textit{Soviet astr.(tr:
Astronomicheskii Zhurnal)}, \textbf{34}, (5), 522.

\bibitem[\protect\citeauthoryear{{Brosius \etal}}{2002}]{Brosius02}
Brosius,~J.W., Landi,~E., Cook,~J.W. \etal: 2002,  \apj,
\textbf{574}, 453.

\bibitem[\protect\citeauthoryear{{Chae}, {Hle} and {Lemaire}}{1998}]{Chae98}
Chae,~J., Hle,~U.S., Lemaire,~P.: 1998, \apj, \textbf{505}, 957.

\bibitem[\protect\citeauthoryear{{Cranmer}}{2004}]{Cranmer04}
Cranmer,~S.R.: 2004, In: R.W.~Walsh, J.~Ireland, D.~Danesy, B.~Fleck
(eds.) \textit{Proceedings of the SOHO 15 Workshop - Coronal Heating
(ESA SP-575)}, Paris: European Space Agency, 154.

\bibitem[\protect\citeauthoryear{{Chiuderi-Drago}}{1974}]{Chiuderi-Drago74}
Chiuderi-Drago,~F., Righini,~G. (ed.): 1974, \textit{Skylab Solar
Workshop: Oss. e Mem. Oss. Arcetri.}, \textbf{164}, 242.

\bibitem[\protect\citeauthoryear{{Chiuderi-Drago \etal}}{1999}]{Chiuderi-Drago99}
Chiuderi-Drago,~F., Landi,~E., Fludra,~A., Kerdraon,~A.: 1999, \ssr,
\textbf{87}(1-2), 141.

\bibitem[\protect\citeauthoryear{{Dulk}, {Sheridan}}{1974}]{Dulk74}
Dulk,~G.A., Sheridan,~K.V.: 1974, \solphys, \textbf{36}, 191.

\bibitem[\protect\citeauthoryear{{Eselevich}}{2009}]{Eselevich09}
Eselevich,~V.G., Fainshtein,~V.G., Rudenko,~G.V. \etal:2009,
\textit{Cosmic Research}, \textbf{47}, 95.

\bibitem[\protect\citeauthoryear{{Grall}, {Coles}, and {Klinglesmith}}{1996}]{Grall96}
Grall,~R.R., Coles,~Wm.A., Klinglesmith,~M.T.:1996, \textit{in Proc.
of the 8 international solar wind conference: Solar wind eight}, AIP
Conf. Proc., \textbf{382}, 108.

\bibitem[\protect\citeauthoryear{{Gopalswamy \etal}}{1999}]{Gopalswamy99}
Gopalswamy,~N., Shibasaki,~K., Thompson,~B.J. \etal: 1999, \jgr,
\textbf{104}(A5), 9767.

\bibitem[\protect\citeauthoryear{{Grechnev \etal}}{2003}]{Grechnev03}
Grechnev,~V.V., Lesovoi,~S.V., Smolkov,~G.Ya. \etal: 2003,
\solphys{}, \textbf{216}(1), 239.

\bibitem[\protect\citeauthoryear{{Gringauz
\etal}}{1962}]{Gringauz62} Gringauz,~K.I., Bezrukikh,~V.V.,
Ozerov,~V.D., Rybchinskii,~R.E.: 1962, \textit{Planetary and Space
Science}, \textbf{9}(3), 103. (First published: 1960,
\textit{Doklady Academy of Sciences U.S.S.R.}, \textbf{131}, 1301)

\bibitem[\protect\citeauthoryear{{Gupta \etal}}{2010}]{Gupta10}
Gupta,~G.R., Banerjee,~D., Teriaca,~L., Imada,~S., Solanki,~S.:
2010, \apj, \textbf{718}, 11.

\bibitem[\protect\citeauthoryear{{Hollweg}}{1978}]{Hollweg78}
Hollweg,~J.V.: 1978, \textit{Rev. Geophis. and Space Phys.},
\textbf{16}, 689.

\bibitem[\protect\citeauthoryear{{Hundhausen}}{1972}]{Hundhausen72}
Hundhausen,~A.J.: 1972, Coronal Expansion and Solar Wind,
Berlin-Heidelberg-New York, Springer-Verlag.

\bibitem[\protect\citeauthoryear{{Kosugi \etal}}{1986}]{Kosugi86}
Kosugi,~T., Ishiguro,~M., Shibasaki,~K.: 1986, \pasj,
\textbf{38}(1), 1.

\bibitem[\protect\citeauthoryear{{Krissinel \etal}}{2000}]{Krissinel00}
Krissinel,~B.B., Kuznetsova,~S.M., Maksimov,~V.P., Prosovetsky,~D.V.
\etal: 2000, \pasj, \textbf{52}, 909.

\bibitem[\protect\citeauthoryear{{Maksimov \etal}}{2004}]{Maksimov04}
Maksimov,~V.P., Prosovetsky,~D.V., Kuznetsova,~S.M., Obukhov,~A.G.:
2004, \textit{Solar-terrestrial physics}, \textbf{6}, 80.

\bibitem[\protect\citeauthoryear{{Maksimov \etal}}{2006}]{Maksimov06}
Maksimov,~V.P., Prosovetsky,~D.V., Grechnev,~V.V., Krissinel,~B.B.,
Shibasaki,~K.: 2006, \pasj, \textbf{58}(1), 1.

\bibitem[\protect\citeauthoryear{{Mercier}, {Klein}, and {Trottet}}{1988}]{MercierNANCAY}
Mercier,~C., Klein,~K.-L., Trottet,~G.: 1988, \adv, \textbf{8}, 193.

\bibitem[\protect\citeauthoryear{{Moran \etal}}{2001}]{Moran01}
Moran,~T., Gopalswamy,~N., Dammasch,~I.E., Wilhelm,~K.A.: 2001,
\aap, \textbf{378}, 1037.

\bibitem[\protect\citeauthoryear{{Nakajima \etal}}{1994}]{Nakajima94}
Nakajima,~H., Nishio,~M., Enome,~S., Shibasaki,~K. \etal: 1994,
\textit{Proc. IEEE}, \textbf{82}(5), 705.

\bibitem[\protect\citeauthoryear{{Neugebauer} and {Snyder}}{1966}]{Neugebauer66}
Neugebauer,~M., Snyder,~C.W.: 1966, \jgr{}, \textbf{71}, 4469.

\bibitem[\protect\citeauthoryear{{Nindos \etal}}{1999}]{Nindos99}
Nindos,~A., Kundu,~M.R., White,~S.M.: 1999, \apj, \textbf{527}(1),
415.

\bibitem[\protect\citeauthoryear{{Obridko \etal}}{2009}]{Obridko09}
Obridko,~V.N., Shelting,~B.D., Livshits,~I.M., Asgarov,~A.B.: 2009,
\solphys{}, \textbf{260}, 191.

\bibitem[\protect\citeauthoryear{{Parker}}{1958}]{Parker58}
Parker,~E.: 1958, \apj{}, \textbf{128}, 664.

\bibitem[\protect\citeauthoryear{{Ponomarev}}{1957}]{Ponomarev57}
Ponomarev,~E.A.: 1957, The theory of solar corona, PhD thesis, Kiev
University publ.

\bibitem[\protect\citeauthoryear{{Sheeley \etal}}{1985}]{Sheeley85}
Sheeley,~N.R.Jr., Howard,~R.A., \etal: 1985, \jgr{}, \textbf{90},
163.

\bibitem[\protect\citeauthoryear{{Shugai \etal}}{2009}]{Shugai09}
Shugai,~Yu.S., Veselovsky,~I.S., Trichtchenko,~L.D.: 2009,
\textit{Geomagnetism and Aeronomy}, 2009, \textbf{49}(4), 41.

\bibitem[\protect\citeauthoryear{{Stepanian \etal}}{2008}]{Stepanian08}
Stepanian,~N.N., Kuzin,~S.V., Fainshtein,~V.G. \etal: 2008,
\textit{Solar System Research}, \textbf{42}(1), 83.

\bibitem[\protect\citeauthoryear{{Stone \etal}}{1998}]{StoneACE98}
Stone, E.C., Frandsen, A.M., Mewaldt, R.A., Christian, E.R.,
Margolies, D., Ormes, J.F., Snow, F.: 1998, \ssr{}, \textbf{86}, 1.

\bibitem[\protect\citeauthoryear{{Trottet}, {Lantos}}{1978}]{Trottet78}
Trottet,~G., Lantos,~P.: 1978, \aap, \textbf{70}, 245.

\bibitem[\protect\citeauthoryear{{Tu \etal}}{1995}]{Tu95}
Tu,~C.-Yi, Zhou,~C., Marsch,~E.: 1995, \ssr, \textbf{73}, 1.

\bibitem[\protect\citeauthoryear{{Tu \etal}}{2005}]{Tu05}
Tu,~C.-Yi; Zhou,~C., Marsch,~E. \etal: 2005, \textit{Science},
\textbf{308}, Issue 5721, 519.

\bibitem[\protect\citeauthoryear{{Vr\v{s}nak \etal}}{2007}]{Vrsnak07}
Vr\v{s}nak,~B., Temmer,~M., Veronig,~A.M.: 2007, \solphys{},
\textbf{240}(2), 315.

\bibitem[\protect\citeauthoryear{{Vsehvyatskiy
\etal}}{1955}]{Vsehvyatskiy55} Vsehvyatskiy,~S.K., Nikol'skiy,~G.M.,
Ponomarev,~E.A., Cherednichenko,~V.I.: 1955,
\textit{Astronomicheskii Zhurnal}, \textbf{32}(2), 165.

\bibitem[\protect\citeauthoryear{{Suzuki}, {Inutsuka}}{2006}]{Suzuki06}
Suzuki,~T., Inutsuka,~S.: 2006, \apj, \textbf{632}(1), L49.

\bibitem[\protect\citeauthoryear{{Wang}, {Sheeley}}{1990}]{Wang90}
Wang,~Y.-M., Sheeley,~N.R.Jr.: 1990, \apj, \textbf{355},726.

\bibitem[\protect\citeauthoryear{{Wang \etal}}{1998}]{Wang98}
Wang,~Y.-M., Sheeley,~N.R., Walters,~Jr.J.H. \etal: 1998, \apj{},
\textbf{498}, L165.

\bibitem[\protect\citeauthoryear{{Wilhelm \etal}}{1995}]{Wilhelm95}
Wilhelm,~K., et al.: 1995, \solphys{} \textbf{162}, 189.


\end{thebibliography}
\end{document}